\documentclass[twocolumn]{aastex63}

\usepackage{graphicx}
\usepackage{amsmath}
\accepted{for publication in ApJ, October 8, 2020}

\shorttitle{Stochastic heating of electrons}
\shortauthors{K. Stasiewicz \& B. Eliasson}

\begin{document}

\title{Stochastic and quasi-adiabatic electron heating in quasi-parallel shocks}


\author[0000-0002-2872-5279]{Krzysztof Stasiewicz}
\email{krzy.stasiewicz@gmail.com}
\affiliation{Department of Physics and Astronomy, University of Zielona G\'ora, Poland }
\affiliation{Space Research Centre, Polish Academy of Sciences, Warsaw, Poland }
\author[0000-0001-6039-1574]{Bengt Eliasson}
\email{bengt.eliasson@strath.ac.uk}
\affiliation{SUPA, Department of Physics, University of Strathclyde, Glasgow, G4 0NG, United Kingdom}

\begin{abstract}
Using Magnetospheric Multiscale (MMS) observations at the Earth's quasi-parallel bow shock we demonstrate that electrons are heated by two different mechanisms:  a quasi-adiabatic heating  process during magnetic field compression, characterized by the isotropic temperature relation $T/B=(T_0/B_0)(B_0/B)^{\alpha}$ with $\alpha=2/3$ when  the electron heating function $|\chi_e|<1$, and a stochastic heating process when $|\chi_e|>1$. Both processes are controlled by the value of 
the stochastic heating function  $\chi_j = m_j q_j^{-1} B^{-2}\mathrm{div}(\mathbf{E}_\perp)$ for particles with mass $m_j$ and charge $q_j$ in the electric and magnetic fields $\mathbf{E}$ and $\mathbf{B}$. Test particle simulations are used to show that the  stochastic electron heating  and acceleration in the studied  shock is accomplished by  waves  at frequencies (0.4 - 5) $f_{ce}$ (electron gyrofrequency) for bulk heating, and waves $f>5\,f_{ce}$ for acceleration of the tail of the distribution function. Stochastic heating can give rise to flat-top electron distribution functions, frequently observed near shocks. It is also shown that obliquely polarized electric fields of electron cyclotron drift (ECD) and ion acoustic instabilities scatter the electrons into the parallel direction and keep the isotropy of the electron distribution.  The results reported in this paper may be relevant to electron heating and acceleration at interplanetary  shocks and other astrophysical shocks.

\end{abstract}



\raggedbottom

\section{Introduction} \label{seq1}

The observational advances enabled by multipoint measurements in space  like Cluster \citep{Escoubet:1997}, THEMIS \citep{Sibeck:2008}, and
MMS \citep{Burch:2016} has stimulated significant progress in space plasma physics. Specifically, the recent MMS mission comprising 4 spacecraft flying through the bow shock and the magnetosheath in formation with separation distances of about 20 km has opened unprecedented possibility for testing theoretical models for heating and acceleration mechanisms that operate at collisionless shocks by detailed comparison with observations. Using these state-of-the-art measurements, which will also be discussed in Section \ref{sec2} of the present paper, \citet{Stasiewicz:2020a,Stasiewicz:2020b,Stasiewicz:2020c} have identified a chain of collective plasma processes that  operate at both quasi-parallel and quasi-perpendicular bow shocks and lead to the heating of ions and electrons. This sequence involves cross-field current driven instabilities and can be summarized as follows:

{\em  Shock compression of the number density $N$ and magnetic field $B$ }    $\rightarrow$ {\em diamagnetic current}  $\rightarrow$ {\em lower hybrid drift (LHD) instability} $\rightarrow$  {\em electron ${\bf E}\times {\bf B}$ drift} $\rightarrow$ {\em modified two stream  (MTS) instability } $\rightarrow$ {\em electron cyclotron drift (ECD) instability}   $\rightarrow$ {\em heating: quasi-adiabatic ($\chi_j<1$), stochastic  ($\chi_j>1$)}.

The above chain of  physical processes is triggered by a single event -- namely -- the  compression of $N$ and $B$ characteristic for fast magnetosonic waves, so the above sequence could  be present in all types of collisionless shock waves in space that are associated with density and magnetic field compression.

The  stochastic heating function of particle species $j$ ($j=e$ for electrons and $p$ for protons) is \citep{Stasiewicz:2020a,Stasiewicz:2020b}

\begin{equation}
\chi_j(t,\mathbf{r})  = \frac{ m_j}{q_j B^2} {\rm div}(\mathbf{E}_\perp),  \label{eq1}
\end{equation}
which is a generalization of the heating condition from earlier works of \citet{Karney:1979,McChesney:1987,Balikhin:1993}, where the divergence is reduced to the directional gradient $\partial E_x/\partial x$. The particles are magnetized (adiabatic) for $|\chi_j|<1$, and demagnetized (subject to non-adiabatic heating) for $|\chi_j|\gtrsim1$.

The high quality electric field measurements by the FIELDS instrumentation suite on MMS \citep{Lindqvist:2016,Ergun:2016,Torbert:2016} makes it possible to directly calculate the divergence of the electric field that is used in Eq.~(\ref{fig1}). The first  time-frequency spectrogram of an approximation to $\mathrm{div}(\mathbf{E})=\rho/\epsilon_0$, i.e., of the electric charge distribution across the bow shock was published by \citet{Stasiewicz:2020a}.

The proton heating function $\chi_p$ typically has values in the range $10-100$ in the bow shock and the magnetosheath, which implies that the ions are strongly demagnetized and can be subject to stochastic heating processes in these regions.

At quasi-perpendicular shocks the  derived values of $\chi_e$ for electrons are mostly below the stochastic threshold, due to the increasing values of $B\approx10-40$ nT in the shock ramp combined with the scaling $\chi_e \propto B^{-2}$. \citet{Stasiewicz:2020c} has shown that such situations instead lead to \emph{quasi-adiabatic electron heating}, characterized by electron heating during the compression of the magnetic field while keeping the magnetic moment conserved. The perpendicular energy gain is redistributed to the parallel energy component by  scattering on waves, leading to the isotropic temperature relation \citep{Stasiewicz:2020c}

\begin{equation}
\frac{T}{B}=\frac{T_0}{B_0}\left(\frac{B_0}{B}\right)^{\alpha} \label{eq2}
\end{equation}
with $\alpha=1/3$, which predicts a dip of $T/B$ where $B$ has a maximum.

The aim of this paper is to make a detailed analysis of electron heating  at quasi-parallel shocks, which complements the work of \citet{Stasiewicz:2020c}, concerned with ion and electron heating at quasi-perpendicular shocks.

\section{Electron heating  in quasi-parallel shocks} \label{sec2}

 \begin{figure}
\includegraphics[width=\columnwidth]{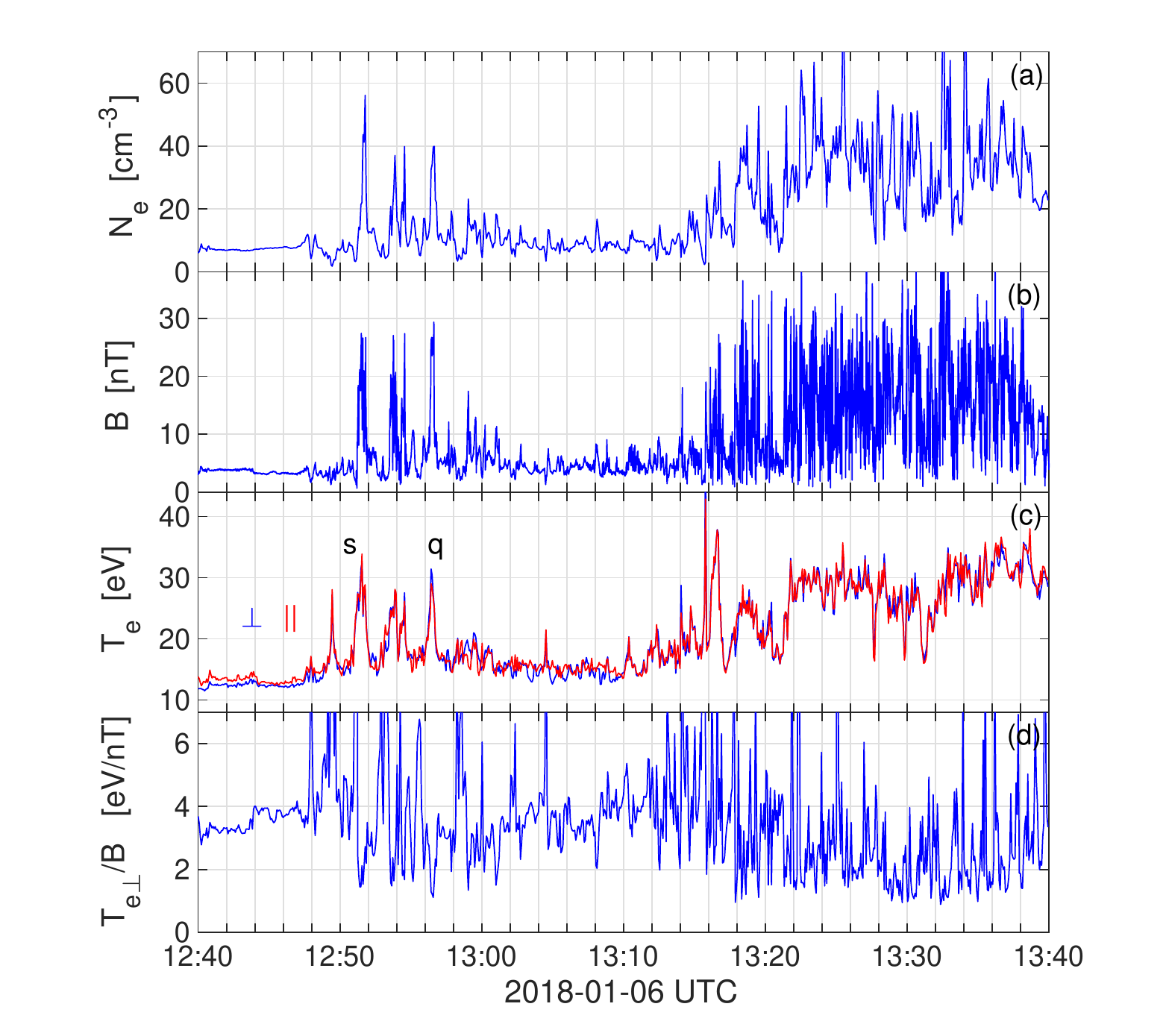}
\caption{MMS-1 measurements from a 1 hr time interval of a quasi-parallel bow shock: (a) electron number density $N_e$ and (b) magnetic field $B$  form large amplitude compressional  structures (shocklets)  -- typical for parallel shocks. (c) Perpendicular and parallel  temperatures of electrons, and  (d) the ratio $T_{e\perp}/B $ that helps to identify the heating processes.  \label{fig1}}
\end{figure}

We analyze  MMS measurements from 2018-01-06 obtained by the 3-axis  electric field sensors \citep{Lindqvist:2016,Ergun:2016,Torbert:2016} and  magnetic field vectors measured by the Fluxgate Magnetometer \citep{Russell:2016},  and the number density, velocity, and temperature of both ions and electrons from the Fast Plasma Investigation  \citep{Pollock:2016}.

When  the interplanetary magnetic field is directed
quasi-parallel to the shock normal, there is usually not a single ramp such as at the perpendicular bow shock, but instead an extended foreshock  region is formed, filled with nonlinear compressional structures (shocklets) like the ones shown in Fig.~\ref{fig1}a,b. These shocklets have spatial scales of $\sim$1,000 km and represent compressions of the plasma density and the magnetic field by a factor 2--10 times  the background values \citep{Schwartz:1991,Stasiewicz:2003,Lucek:2008,Wilson:2013}. The large amplitude shocklets are typically standing against the solar wind flow and move with speeds of the order tens km/s with respect to the spacecraft. A frequently used an awkward acronym for these structures is SLAMS (short large amplitude magnetic structures), while the term 'shocklets' is sometimes misleadingly  used for long-wavelength $\sim$R$_E$, compressional magnetosonic waves.

The large amplitude compressions of $N$ and $B$ seen in Fig.~\ref{fig1}a,b are likely to trigger the chain of cross-field current driven instabilities that can lead to the heating and acceleration of ions and electrons, as mentioned in the Introduction.
Indeed, in Fig.~\ref{fig1}c we see localized enhancements of the electron temperature, related to compressions of $N$ and $B$, with almost equal perpendicular and parallel temperatures, $T_{e\perp}\approx T_{e\parallel}$. The ratio $T_{e\perp}/B$ shown in Fig.~\ref{fig1}d has been found to be a good indicator of the heating processes \citep{Stasiewicz:2020c}. A flat ratio would indicate adiabatic process related to the conservation of the magnetic moment, but this is generally not observed. A  dip in the ratio is characteristic to quasi-adiabatic processes described by Eq.~(\ref{eq2}).
A hump in the ratio may indicate non-adiabatic (possibly stochastic) heating, but may also be caused by other reasons as for example a magnetic depression not associated with heating, or a passage of a plasma cloud heated  at an earlier time. Visible in Fig.~\ref{fig1}d are many humps and smaller amplitude dips of $T_{e\perp}/B$, which indicates that both stochastic and quasi-adiabatic heating may be operational. The event marked 's' in Fig.~\ref{fig1}c is a representative for stochastic heating, and the event 'q' for quasi-adiabatic heating. In these two events the electron temperature is increased by $\sim$100\%, but there are many smaller heating events, with a few percent increase of the electron temperature throughout the whole time interval.

\begin{figure}
\includegraphics[width=\columnwidth]{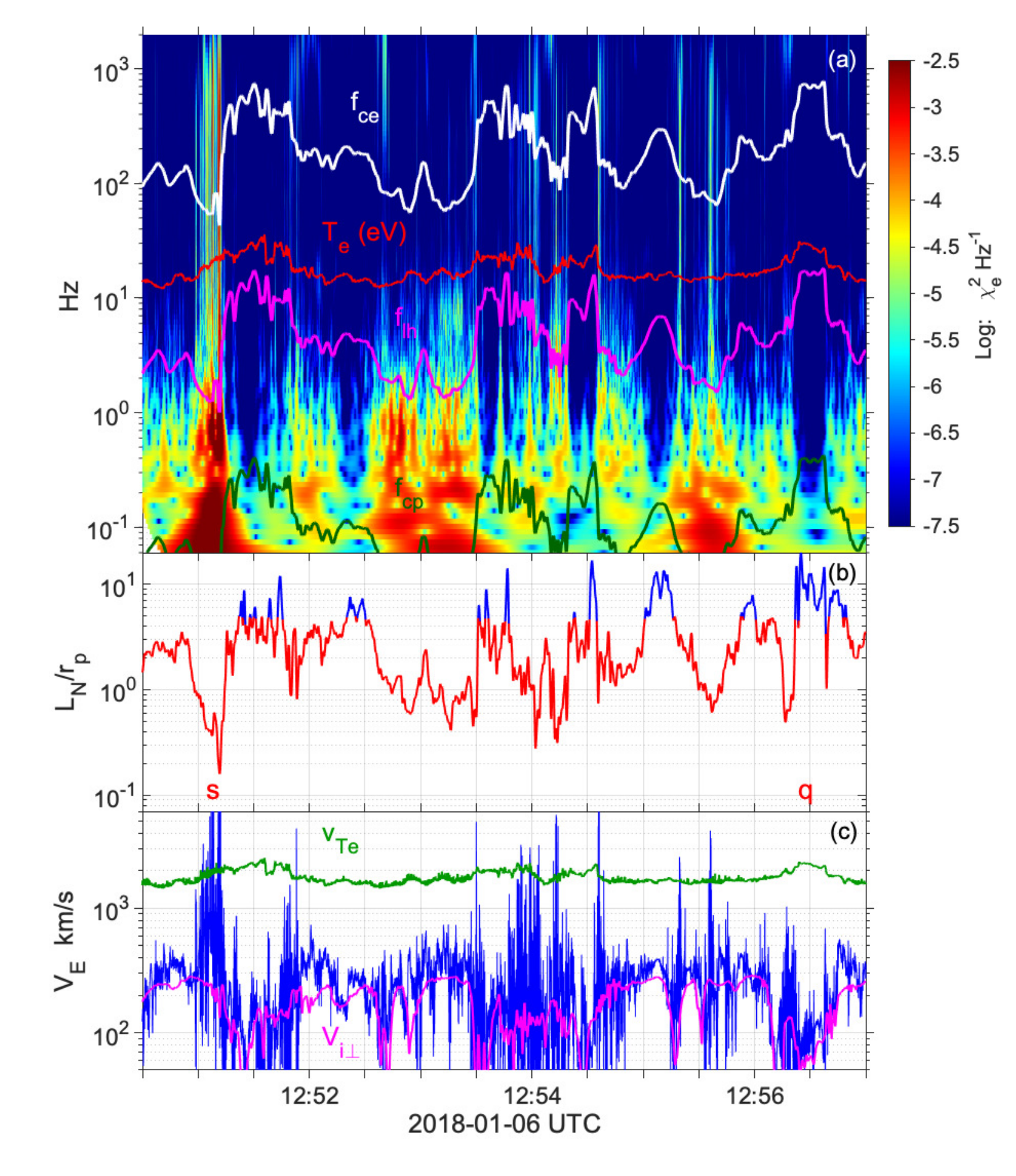}
\caption{(a) Time vs frequency spectrogram of $\chi_e$ for the time interval 12:50:30-12:57:00 UTS of Fig. \ref{fig1}. Over-plotted are the proton cyclotron frequency $f_{cp}$, the lower hybrid frequency $ f_{lh}$, the electron cyclotron frequency $ f_{ce}$, and the electron temperature $T_{e\perp}$ (eV, numerical values for the left scale apply).  (b) The gradient scale of the plasma density $L_N$  normalized with thermal proton gyroradius $r_p$ (median = 200 km). Regions with  $L_N/r_p \lesssim 5$ are  unstable for the LHD instability (marked red). Markers for events 's' and 'q' of Fig. \ref{fig1} are shown in the bottom.  (c) The computed ${\bf E}\times {\bf B}$ drift speed $V_E$ (blue), the electron thermal speed $v_{Te}$ (green), and the ion perpendicular speed $V_{i\perp}$ (magenta). Waves between $f_{cp}$ and $f_{lh}$  are attributed to the LHD and above $f_{lh}$ to the MTS (modified two-stream) instabilities, and for frequencies around $f_{ce}$ and above to the ECD instability. Note the vertical striations that start from below 1 Hz (LHD instability) and go through the MTS and ECD instabilities up to 2 kHz, indicating co-location and common origin of these instabilities.   \label{fig2}}
\end{figure}

In Fig.~\ref{fig2} we make a detailed analysis of data for the time interval 12:50:30 to 12:57:00 UTC, which includes the 's' and 'q' events. The median values for the plasma parameters during this time interval are: the plasma  beta: $\beta_e\sim 2,\, \beta_i\sim 18$, the sound Mach number 1.5 and the Alfv\'en Mach number 7, the electron gyrofrequency $f_{ce} \approx 154$ Hz, the thermal proton and electron  gyroradii $r_p\approx 200$ km and $r_e\approx 2$ km, respectively, the perpendicular ion flow velocity is $V_{i\perp}\approx 200$ km/s, and the electron and ion temperatures 20 and 120 eV, respectively. The position of these observations was (12.7, 6.9, 4.7) R$_E$ GSE, and the average  inter-spacecraft distance was 25 km.

In Figure \ref{fig2}a we show time vs frequency spectrogram of $\chi_e$ with ${\rm div}(\mathbf{E}_\perp)$ computed from 4-point measurements using the method of  \citet{Harvey:1998} developed for Cluster. Over-plotted are: the proton cyclotron frequency $f_{cp}=(eB/m_p)/2\pi$, the lower hybrid frequency $ f_{lh}=(f_{cp}f_{ce})^{1/2}$,  the electron cyclotron frequency $f_{ce}=(eB/m_e)/2\pi$, and the electron temperature $T_{e\perp}$ (eV).

 There are concerns \citep{Goodrich:2018} that the axial double probe (ADP)  on MMS, which uses  rigid axial booms shorter than the wire booms of the spin-plane double probe experiment (SDP), produces larger amplitude responses for short, tens of meter (Debye length) waves like ion-acoustic (IA) waves. This instrumental asymmetry may be propagated and affect the computations of gradients of the electric field and the corresponding value of $\chi_e$. For this reason, the computations of  div(\textbf{E}) are made in the despun spacecraft coordinates (DSL), which separate $E_z$ provided by the ADP from ($E_x,E_y$) provided by the SDP. In these coordinates it is possible to remove the highest frequency components above $f_{ce}$ from the analysis,   which may contain such short waves around the proton plasma frequency $f_{pp}$, or to not use the $E_z$ component at all. For the purpose of the spectrogram shown in Fig. \ref{fig2}a we applied rescaling of the $E_z$ component  to assure  that $\mathrm{rms}(E_z)=\mathrm{rms}(E_y)$.

Figure \ref{fig2}b shows the gradient scale length $L_N= N|\nabla N|^{-1}$ for the electron density $N$ derived from  data using the method of  \citet{Harvey:1998}.
When the  scale of the density gradient obeys the condition $L_N/r_p<(m_p/m_e)^{1/4}$ the ion diamagnetic drift $V_{di}=T_p(m_p\omega_{cp}L_N)^{-1}=v_{Tp}(r_p/L_N)$ exceeds the threshold for the onset of the  LHD instability  \citep{Davidson:1977,Drake:1983,Gary:1993}. Here, $v_{Tp}=(T_{p\perp}/m_p)^{1/2}$ is the proton thermal speed and $r_p=v_{Tp}/\omega_{cp}$ the proton Larmor radius.

Waves  between $f_{cp}$ and   $f_{lh}$ in panel (a) are  related to the LHD instability, which has maximum growth rate at $k_\perp r_e\sim 1$  \citep{Davidson:1977}, however simulation results of \citet{Daughton:2003} indicate that they have longer wavelengths $k_\perp(r_e r_p)^{1/2}\sim 1$ and electromagnetic character with significant magnetic fluctuations, which are observed also in the present case. Here, $k_\perp$ is the wavenumber perpendicular to the magnetic field, and $r_e=v_{Te}/\omega_{ce}$ is the electron thermal Larmor radius, and $v_{Te}=(T_{e\perp}/m_e)^{1/2}$ is the electron thermal speed.

The enhanced electric field of the LHD waves produces strong ${\bf E} \times {\bf B}$ drifts of electrons only, because the ions are not subject to this drift due to their large gyroradius in comparison to the width of drift channels. When the electron-ion  drift exceeds the ion thermal velocity, the modified two-stream  (MTS) instability can also be excited, which may be responsible for the waves observed with frequencies above $f_{lh}$ \citep{Lashmore:1973,Gary:1987,Umeda:2014}. The MTS and LHD instabilities belong to the same dispersion surface \citep{Silveira:2002,Yoon:2004}, so we will not make distinction between these instabilities and use term LHD instability in the sense of a generalized cross-field current driven instability in the lower-hybrid frequency range.

When the relative electron-ion drift speed becomes a significant fraction of the electron thermal speed, $V_{E}=|{\bf E}\times{\bf B}|/B^2\sim v_{Te}$, the ECD instability is initiated, which creates even larger electric fields on spatial scales of $r_e$ and smaller \citep{Forslund:1972,Lashmore:1971,Muschietti:2013,Janhunen:2018}.

This process can be inferred from Fig.~\ref{fig2}c which shows the electric drift speed $V_E$ computed from the measured electric field in the  frequency range 0 -- 256 Hz (blue curve). The highest frequencies were removed because at $f>f_{ce}$ the electron drift approximation is not valid. For comparison we plot also the electron thermal speed $v_{Te}$ and the measured perpendicular speed of the ions $V_{i\perp}$ (magenta).  Large values  of $V_E\sim v_{Te}$ are signatures of large electric fields ($E\sim 100$ mV/m) of  ECD waves which may be  Doppler downshifted and observed also below $f_{ce}$. Large differences between the electron drift $V_E$ and the measured perpendicular drift of ions $V_{i\perp}$ would induce sequentially the MTS and ECD instabilities after initiation of the LHD instability on the density gradients. The high $V_E$ drift regions in Fig. \ref{fig2}c are manifestations of spatially coupled LHD, MTS, and ECD instabilities.

The  ECD instability is driven by the cyclotron resonance $\omega - k_\perp V_{de}= n\omega_{ce}$, where  $n$ is an integer, and  $V_{de}\approx V_E$ is the electron drift velocity in the rest frame of ions \citep{Janhunen:2018}. For $\omega$ near cyclotron harmonics this resonance condition is $k_\perp V_{E} \approx n\omega_{ce}$, which can be expressed equivalently by

\begin{equation}
k_\perp r_e \approx \frac{nv_{Te}}{V_E }\label{kr}
\end{equation}
 which  implies that the  wavelength is

 \begin{equation}
 \lambda \approx 12.6\,[\mathrm{km}]\, \frac{V_E}{nv_{Te}} , \label{lambda}
 \end{equation}
with the numerical coefficient  given for $r_e=2$ km.
This  means that ECD waves with $n=1$ and electric drift velocities $V_E>v_{Te}$  have wavelengths that enable accurate gradient computations needed for the calculations of div($\mathbf{E}$) by the MMS spacecraft constellation.
 Because of short wavelengths, the higher harmonics $n>1$ may be Doppler  downshifted by $\sim 100$ Hz and observed in the $f_{lh}-f_{ce}$ frequency range. Their contribution to $\chi_e$ will be underestimated by the gradient computation procedure.

It should be emphasized that derivatives $\partial E_x/\partial x, \, \partial E_y/\partial y,\, \partial E_z/\partial z$ needed for the $\mathrm{div}(\mathbf{E})$ computation depend on the partial separations of the satellites and not on the radial distances. The partial separations can be much smaller than the radial ones. For example, in the analyzed case we have $\Delta x_{14}=0.3$ km, $\Delta y_{34}=4$ km, $\Delta z_{13}=0.8$ km. In case of waves propagating in the $x-$direction, satellites '1' - '4' could accurately capture gradients on  1 km scale, while satellites '1' - '3' would capture correctly $\sim 1$ km wave propagating in the $z-$direction. This is under assumption that the transverse scale of the wave front is larger than the tetrahedron.  For an arbitrary wave  one should find similar values for projections of the relative vector  ($\mathbf{r}_i -\mathbf{r}_j$) on the wave propagation direction. Thus, the accuracy of the gradient determination  by the \citet{Harvey:1998} method is generally better than the nominal radial distances between the satellites  ($\sim$ 20 km) would imply.

Waves around $f_{ce}$ and above that are associated with the ECD instability have been observed at the bowshock as discussed by \citet{Wilson:2010,Breneman:2013,Goodrich:2018} and \citet{Stasiewicz:2020a}.  These waves have both a perpendicular component of the electric field and a comparable parallel $E_\parallel$ component, related to ion-acoustic (IA) waves \citep{Fuselier:1984,Mozer:2013,Goodrich:2018}.

 \subsection{Quasi-adiabatic heating: event 'q'}

\begin{figure}
\includegraphics[width=\columnwidth]{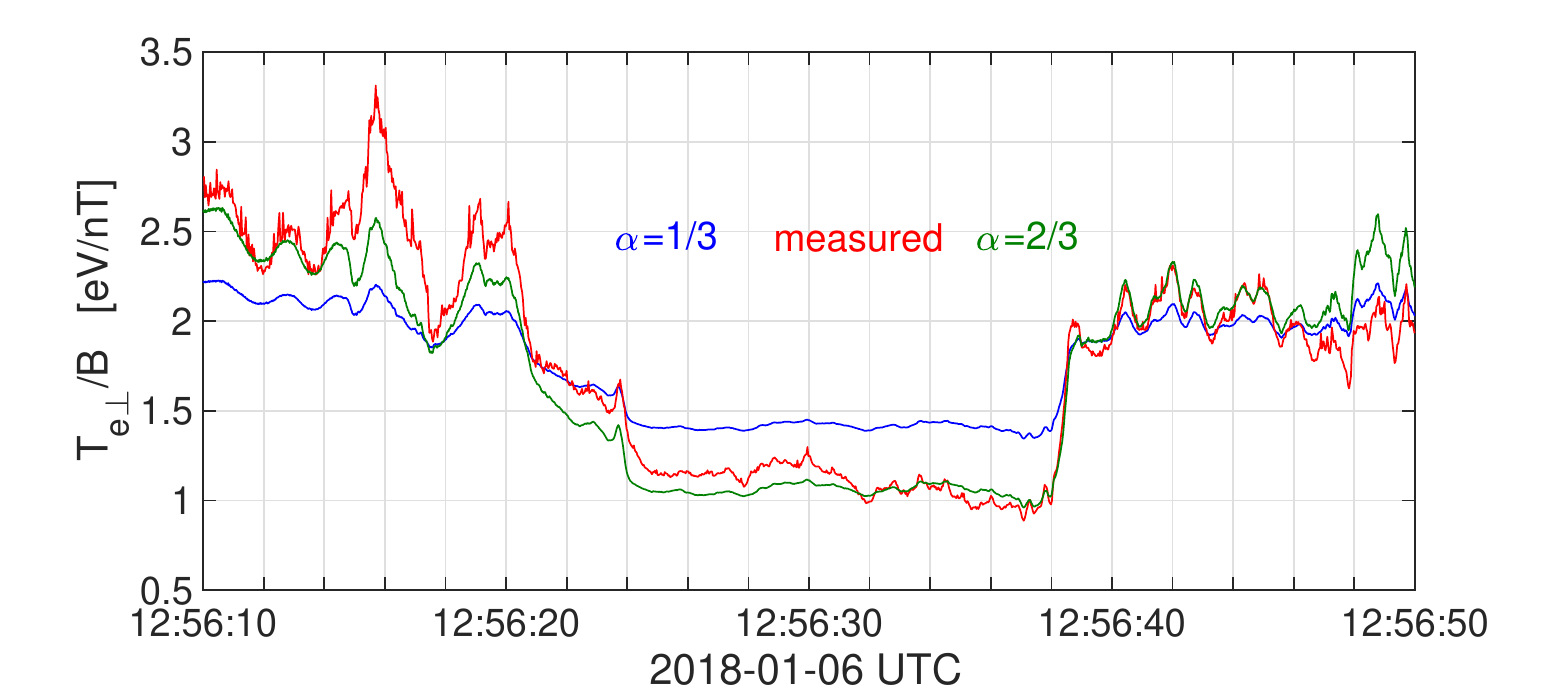}
\caption{ Comparison of the measured ratio $T_{e\perp}/B$ with Eq. (\ref{eq2}) for the event labeled 'q' in Fig.~\ref{fig2}d. The measured curve is in red, modeled for $\alpha=1/3$ in blue, and modeled for $\alpha=2/3$ in green. The latter gives a better fit with observations as further explained in the text. \label{fig3}}
\end{figure}

The quasi-adiabatic event 'q' in Fig.~\ref{fig1}c is clearly related to the compression of $B$ shown in panel (b), and a dip in the $T_{e\perp}/B$ ratio shown in panel (d).
Figure \ref{fig3} shows a detailed  comparison between the measured ratio in red color and the theoretical expression (\ref{eq2})  with $\alpha=1/3$ in blue. We also show the third curve in green, computed with $\alpha=2/3$, which fits the observations significantly better than with $\alpha=1/3$.

Equation (\ref{eq2}) was derived by \citet{Stasiewicz:2020c} with the following justification: When the magnetic moment is conserved, i.e., $T_\perp/B=const$, the differential temperature increase is $dT_\perp=T_\perp B^{-1}dB$. If the energy gain from 2 perpendicular degrees of freedom ($2dT_\perp$) is redistributed by pitch angle scattering to 3 degrees of freedom ($3dT$)  the conservation of energy implies

\begin{equation}
3dT=2TB^{-1}dB
\label{eq3}
\end{equation}
for $T=T_\perp=T_\parallel$. The solution of this equation is given by (\ref{eq2}) with $\alpha=1/3$, and provides a very good approximation to the ratio measured in perpendicular shocks \citep{Stasiewicz:2020c}.
The fit is not as good in the present case. Let us make a reasonable assumption that there are additional sinks of energy, and  an amount   $\delta W=3dT$ goes for the production of waves, in addition to $3dT$ for the isotropization. The modified energy equation is

\begin{equation}
6dT=2TB^{-1}dB,
\label{eq4}
\end{equation}
which has the solution (\ref{eq2}) with exponent $\alpha=2/3$.

A plausible explanation for the difference between the quasi-adiabatic response in quasi-perpendicular shocks ($\alpha=1/3$) and in the present case of quasi-parallel shocks ($\alpha=2/3$) is that in the previous case the waves engaged with the isotropization were generated by ${\bf E}\times{\bf B}$ electron drift caused by lower frequency waves, so the whole energy gain from adiabatic perpendicular heating was used for isotropization. In the present case, the waves performing isotropization are fed by the adiabatic build-up of the temperature anisotropy. There are high-frequency electrostatic waves and  whistlers $\sim$100 Hz associated with this case, which may provide the required sink of kinetic energy.
There may be other redistributions of the energy in other cases, and the values of $\alpha$ may be found different from $\alpha=1/3,2/3$.
 The correspondence of the two curves in Fig.~\ref{fig3} indicates outstanding quality of particle measurements by the FPI instrument \citep{Pollock:2016}, which is able to reproduce the subtle effects of Eq.~(\ref{eq2}) in sharp gradients of the shocklet in event 'q'.

\subsection{Observations of the stochastic heating}

In Figure \ref{fig2}a  waves between $f_{cp}-f_{lh}$  are attributed to the LHD and above $f_{lh}$ to the MTS (modified two-stream) instabilities, and for frequencies around $f_{ce}$ and above to the ECD instability. Note the vertical striations that start from below 1 Hz (LHD instability) and go through the MTS and ECD instabilities up to 2 kHz, indicating co-location and common origin of these instabilities. 
The profile of the magnetic field can be seen in plots of the plasma frequencies $f_{cp},\, f_{lh},\,f_{ce}$, which are all proportional to $B$, and directly in Fig. \ref{fig1}b.
The stochastic heating event 's'  occurs in a region with strong density gradients in panel (b), at the foot of a large amplitude shocklet seen in the frequency plots. It is associated with $\chi_e\sim 10$  and strong spectrum, which rapidly quenches just after 12:51:15 when  the magnetic field increases. After this time the heating mostly continues as compressional, quasi-adiabatic with $|\chi_e|<1$.

 It is remarkable, that the measured density gradient in panel (b) appears to obey the instability condition derived some 40 years ago \citep{Davidson:1977,Drake:1983}, $L_N/r_p<(m_p/m_e)^{1/4}\approx 6$, for the onset of the  LHD instability. The LHD waves are seen in panel (a) below the lower hybrid frequency. The unstable regions are marked red in panel (b).

The data presented in Fig.~\ref{fig2}a-c show correlations between the density gradients (panel b) and strong $V_E$ drifts induced by  large electric fields of LHD/ECD waves (panel c). The $V_E$ values, determining the drift of electrons are much larger than the measured drift of ions $V_{i\perp}$, which would trigger consecutively  the LHD/MTS and ECD instabilities that have different thresholds.

 All these signatures  support the suggested heating scenario at shocks, which starts with the compression of $N$ and $B$, then develops LHD/MTS/ECD instabilities on the gradients and induced drift velocities, and further lead to either quasi-adiabatic or stochastic heating controlled by the stochastic function $\chi_j$ \citep{Stasiewicz:2020a,Stasiewicz:2020b,Stasiewicz:2020c}.

Plasma structures with oppositely directed currents and turbulence characteristic for quasi-parallel shocks create minima of $B$, or even $B\sim 0$ regions (see Fig. \ref{fig1}b), so that regions with $|\chi_e|>1$ and associated stochastic heating is more common in parallel shocks than in perpendicular ones.

In this paper we focus on the electron heating, but it is worth to mention that in case of Figure \ref{fig2} we observe spectacular  stochastic heating of ions. Protons are rapidly heated  from 40 eV to 500 eV with suprathermal tails accelerated to $\sim 10$ keV in three bursts related to the enhancements of the lower hybrid drift waves seen in panel (a). The stochastic ion bulk heating by LHD waves ($f_{cp}<f <f_{lh}$) and the tail acceleration mechanism by LH(MTS) waves ($f\gtrsim f_{lh}$) is universal,  the same as in quasi-perpendicular shocks analyzed by \citet{Stasiewicz:2020c}.

To assess which wave frequencies and wavelengths from a broad spectrum of Figure \ref{fig2} contribute most to the stochastic heating and isotropization of electrons we perform test particle simulations described in the next section.

\section{Simulations of stochastic heating and isotropization} \label{sec3}

We follow the simulation setup described in detail by \citet{Stasiewicz:2020c}, extended here by including a magnetic-field-aligned component of the electrostatic wave field. The parallel component of the wave electric field is essential for isotropization processes.
We consider a spacecraft/observer coordinate system in which the magnetic field $B_0$ is in the $z$ direction, and a macroscopic convection electric field $E_y=E_{0y}$ drives particles into an electrostatic wave with electric field $(E_x,E_z)=(E_{0x},\,E_{0z})\cos(k_x x + k_z z -\omega_s t)$ with frequency $\omega_s$ in the spacecraft frame, 
propagating  in the $x$-$z$ plane, at angle $\theta$ to the magnetic field, where $\tan(\theta)=E_x/E_z=k_x/k_z$. We keep the magnetic field constant to separate purely stochastic heating from the quasi-adiabatic heating in compressed magnetic fields. Thus, in the Doppler-shifted frame of the satellite, the drifting plasma is characterized by the convecting electric field and the time-dependent wave electric field. The governing equations for particles with mass $m$, charge $q$ are

\begin{figure*}
\includegraphics[width=18cm]{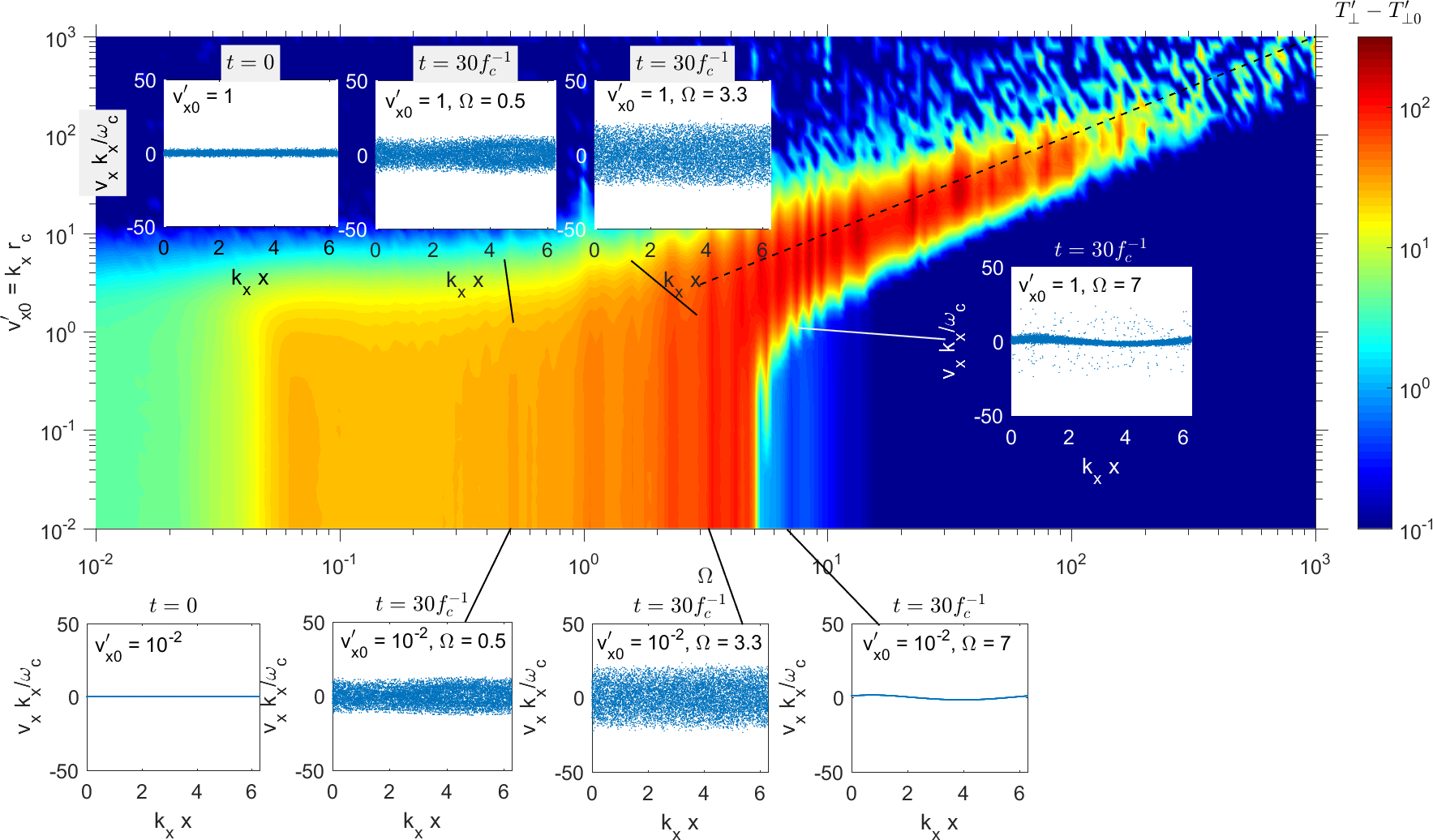}
\caption{A color plot of stochastic heating showing the difference between the final and initial temperatures, $T_\perp^\prime-T_{\perp0}^\prime$,  after 30 cyclotron periods, for charged particles in an electrostatic wave with normalized electric field amplitude $\chi=10$ directed perpendicularly to the magnetic field, $\theta=\pi/2$. Here $f_c=\omega_c/2\pi$,  $T_{\perp0}^\prime = v_{\perp0}'^2$ is the normalized initial temperature and $v_{x0}'=k_x v_{T_{\perp0}}/\omega_c$ with the thermal speed $v_{T_{\perp0}}=(T_{\perp0}/m)^{1/2}$. The insets show distribution functions in ($x$, $v_x$) space at $t=0$ and $30\,f_{c}^{-1}$ for different values of $\Omega$ and $v_{x0}'$. Bulk heating takes place for $\Omega\lesssim 5$, while for $\Omega\gtrsim 5$ there is significant heating only for thermal velocity comparable to the phase velocity, or $v_{x0}'\sim\Omega$ in the normalized variables (dashed line) leading to a distribution function having a high energy tail of particles.  \label{fig4}}
\end{figure*}

\begin{align}
   m \frac{dv_x}{dt}&=q E_{0x}\cos(k_x x + k_z z -\omega_s t)+q v_y B_0,\label{eq_D1}\\
   m \frac{dv_y}{dt}&=qE_{0y}-q v_x B_0,\\
   m \frac{dv_z}{dt}&=q E_{0z}\cos(k_x x + k_z z -\omega_s t),\\
  \frac{d x}{dt}&=v_x,\quad
  \frac{d y}{dt}=v_y,\quad
  \frac{d z}{dt}=v_z.
   \label{eq_D2}
\end{align}

By a change of frame into that of the ${\bf E}\times {\bf B}$-drift velocity (the plasma frame),

\begin{equation}
  v_x=V_x+\frac{E_{0y}}{B_0}, \quad
   x=X+\frac{E_{0y}}{B_0}t,
\end{equation}
and the wave frequency in the plasma frame becomes $\omega=\omega_s -k_x E_{0y}/B_0$,  incorporating the electric drift velocity. Using dimensionless variables with time normalized by $\omega_c^{-1}$, space by $k_x^{-1}$ and velocity by $\omega_c/k_x$ with $\omega_c=q B_0/m$ being the angular cyclotron frequency, gives the system of dimensionless, primed variables,

\begin{align}
  \frac{dv_x'}{dt'}&= \chi \cos( x'+\kappa z' - \Omega t')+ v_y',\label{eq_normed1} \\
  \frac{dv_y'}{dt'}&= -v_x',\\
  \frac{dv_z'}{dt'}&= \chi \kappa \cos( x'+\kappa z' - \Omega t'),\label{eq_normed1_2}\\
  \frac{d x'}{dt'}&=v_x',\quad
  \frac{d y'}{dt'}=v_y',\quad
  \frac{d z'}{dt'}=v_z',\label{eq_normed2}
\end{align}
where $\kappa=k_z/k_x=E_{0z}/E_{0x}=1/\tan(\theta)$ describes the direction of the electrostatic wave to the magnetic field,

\begin{equation}
\Omega =\omega_c^{-1}(\omega_s -k_x E_{0y}/B_0)  \label{eq_om}
\end{equation}
is the normalized wave frequency in the plasma frame, and

\begin{equation}
  \chi  =\frac{k_x}{\omega_c}\frac{E_{0x}}{B_0}
  \label{eq_chi_perp}
\end{equation}
is the stochastic heating parameter, equivalent to Eq.~(\ref{eq1}), representing the normalized perpendicular component of the wave amplitude.
The normalized parallel component of the wave amplitude $\chi\kappa$ in Eq.~(\ref{eq_normed1_2}) is found in Section \ref{seq_iso} to be
responsible for the isotropization of the distribution function.
Stochastic motion takes place only in restricted regions in phase space \citep{Karney:1979,Fukuyama:1977,McChesney:1987}, and hence the initial velocity may also be considered a parameter. For a statistical description of the particles, the initial condition is taken to be a bi-Maxwellian distribution function, which in normalized variables takes the form

\begin{equation}
  F'=\frac{1}{(2\pi)^{3/2} v_{x0}'^3\tau^{1/2}}\exp\bigg(-\frac{(v_x'^2+v_y'^2+v_z'^2/\tau)}{2 v_{x0}'^2}\bigg)
\end{equation}
where $v_{x0}'=k_x r_c$ is the normalized thermal speed, $r_c=v_{T_{\perp0}}/\omega_c$ is the thermal Larmor radius, and $\tau=T_{\parallel0}/T_{\perp0}$ is the initial parallel-to-perpendicular temperature ratio; we will use $\tau=0.01$ below. Here $v_{T_{\perp0}}=(T_{\perp0}/m)^{1/2}$ is the initial thermal speed and $T_{\perp0}$ is the initial perpendicular temperature.
The value of $v_{x0}^{\prime}$ determines the initial temperature in the velocity distribution function, which due to the normalization, is in fact proportional to the ratio of the gyroradius to the wavelength $\lambda=2\pi/k_x$.
The perpendicular kinetic temperature resulting from the stochastic heating is calculated as

\begin{equation}
  T_\perp=m(\langle v_x^2\rangle  -\langle v_x\rangle^2 + \langle v_y^2\rangle  - \langle v_y\rangle^2 )/2,
\end{equation}
while the parallel temperature is obtained as

\begin{equation}
  T_\parallel=m(\langle v_z^2\rangle -\langle v_z\rangle^2),
\end{equation}
and the total kinetic temperature $T=(2T_\perp+T_{\parallel})/3$. Here the angular brackets denote averages over particles, $\langle u \rangle = (\sum_{k=1}^M u_k)/M$.
The system (\ref{eq_normed1})--(\ref{eq_normed2}) is advanced in time using a St{\"o}rmer-Verlet scheme \citep{Press:2007}.

\subsection{Simulations of the perpendicular heating} \label{seq_sim}

The above derived simulation equations will first be used to study the perpendicular heating in parameter space ($\Omega,\,k_x r_c,\,\chi$), so we set $\theta=90^\circ$. We carry out a set of test particle simulations for  $M=10\,000$ particles, which are Maxwell distributed in velocity and uniformly distributed in space.  The input variables for the simulations are: the normalized wave frequency $\Omega$ in the range $10^{-2}$ to $10^3$, and the initial normalized thermal velocity  $v_{x0}'=k_x r_c$ spanning $10^{-2}$ to $10^3$. The normalized amplitude of the electrostatic wave is set to $\chi=10$, consistent with the 's' event in Fig.~\ref{fig2}d.

Simulations are carried out for different values of $\Omega$ and $v_{x0}'$ to produce the color plot in Fig.~\ref{fig4}, which shows the difference $T_\perp^\prime-T_{\perp0}^\prime$ between the normalized kinetic temperature $T_\perp^\prime=k_x^2 T_\perp/m\omega_c^2$ at the end of the simulation and the initial value $T_{\perp0}^\prime=(v_{x0}')^2=k_x^2 T_{\perp0}/m\omega_c^2$.

The simulations are run for 30 cyclotron periods of the particles to primarily study electron heating, while in the previous case \citep{Stasiewicz:2020c} focused on ion heating the time was 3 periods, and $\chi=60$. However, the colormaps in Fig.~\ref{fig4} and in the previous paper  are applicable for any charged particle species. Comparing these two maps we see that the limiting upper frequency for the bulk heating is determined by the value of $\chi$. In this case for $\chi=10$ it is $\Omega\approx 5$, while for $\chi=60$ this boundary moves up to $\Omega\approx 10$. On the other hand, the lower frequency of the bulk heating region extends to lower values with increasing interaction time. It was at $\Omega\approx 1$ after 3 cyclotron periods for $\chi=60$, and moved down to $\Omega\approx 0.05$ after 30 cyclotron periods in the present case.

Strong electron heating is seen for frequencies above $f_{ce}$, but a weak heating region extends down to lower hybrid frequencies $\Omega\sim 0.03$. Waves at lower frequencies  have longer wavelengths and the electric field amplitude usually much smaller than the ECD waves, so they would produce typically $\chi_e <1$. Thus, the $\sim f_{lh}$ frequency waves should not be capable to heat perpendicularly the electron population.  

\begin{figure}
\includegraphics[width=\columnwidth]{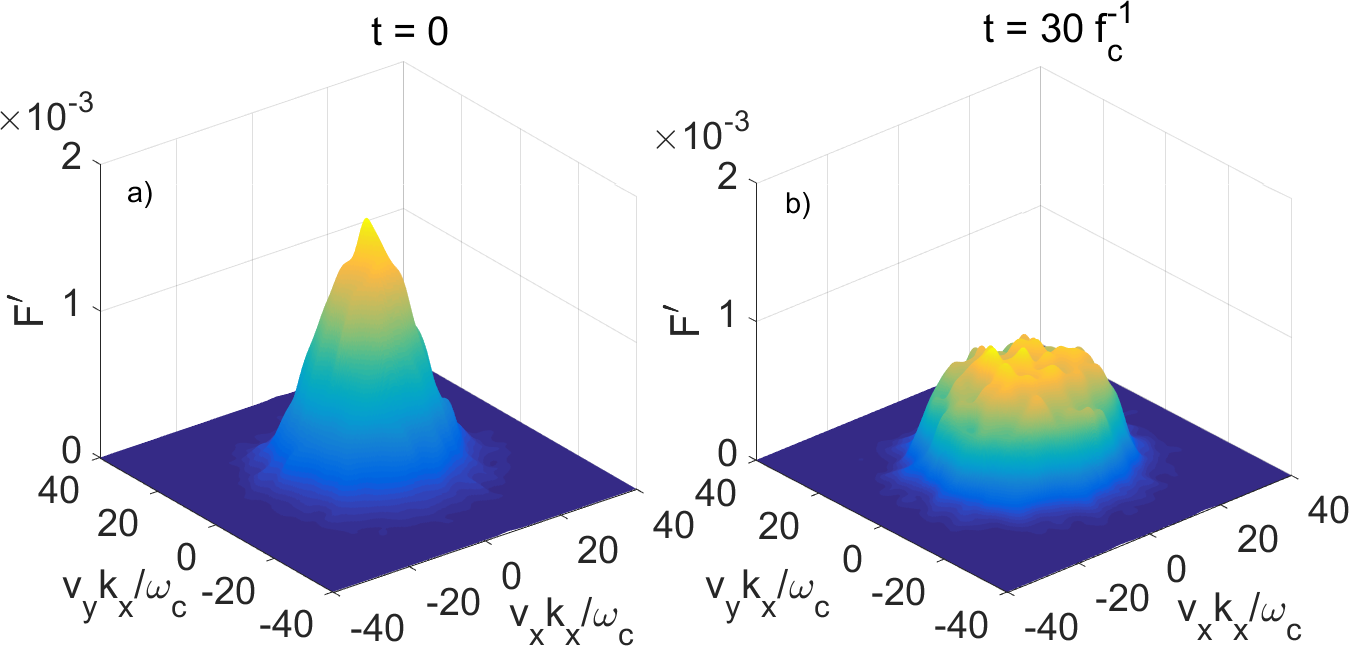}
\caption{The formation of a flat-top electron distribution for $\chi=10$, $\Omega=3.3$ and $v_{x0}^\prime=10$.  The initial Maxwell distribution (a) at $t=0$ becomes the flat-top distribution (b)  after $t=30\,f_c^{-1}$. \label{fig_flattop}}
\end{figure}

To visualize the effect of the stochastic mechanism on the particle distribution function we  perform a test-particle simulation with $\chi=10$, $\Omega=3.3$ and $v_{x0}^\prime=10$ shown in Figure \ref{fig_flattop}. We observe that  the initially Maxwellian distribution function forms a flat-top distribution  after 30 electron cyclotron periods. Flat-top electron distributions are commonly observed at shocks \citep{Feldman:1983,Lefebvre:2007}, and these observations can be explained by the stochastic mechanism discussed in this paper.

\subsubsection{Stochastic heating: event 's'}
\begin{figure}
\includegraphics[width=\columnwidth]{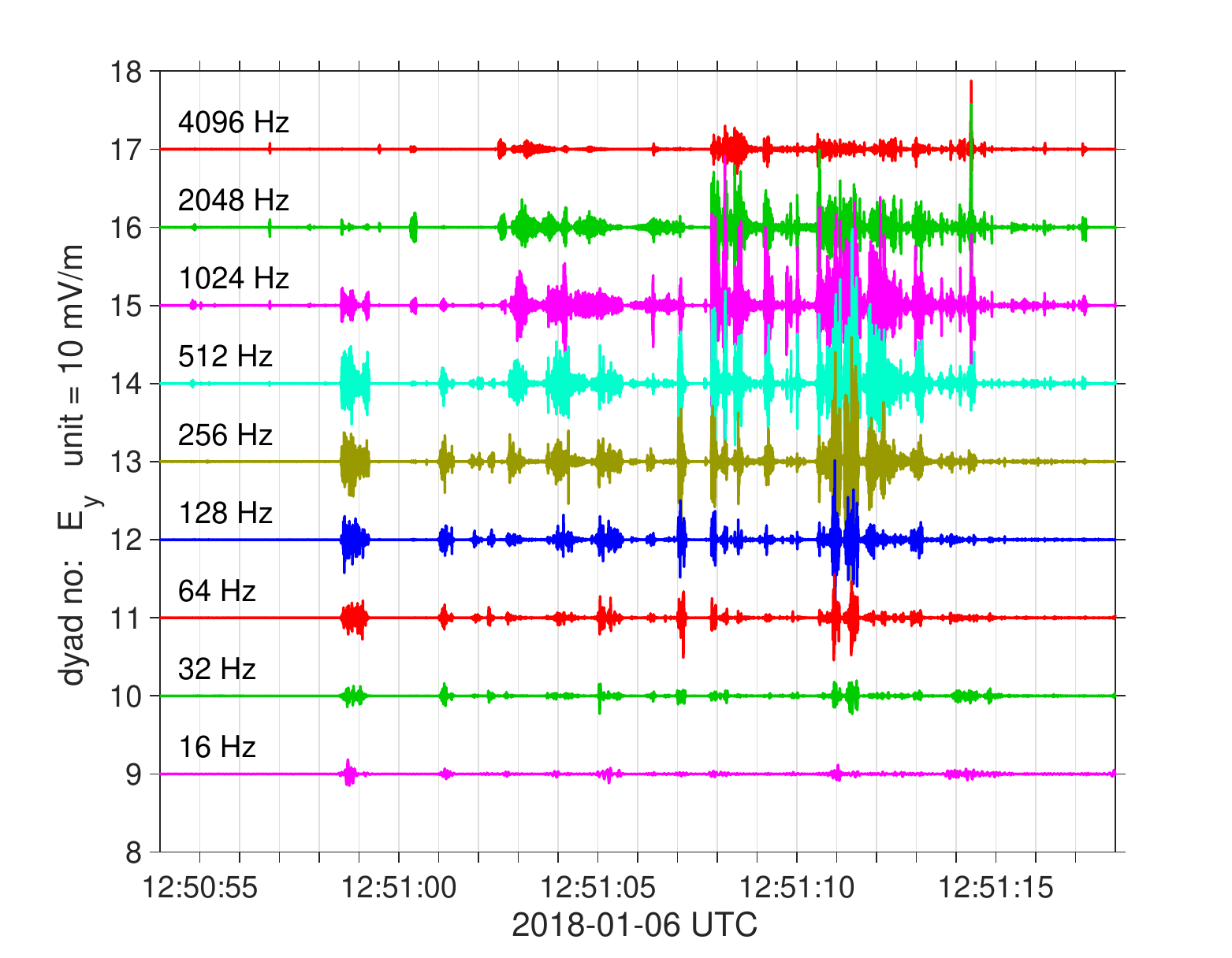}
\caption{ Decomposition of the measured electric signal $E_y$ of waves during the stochastic event 's' into discrete frequency dyads using orthogonal wavelets. Unit amplitude is 10 mV/m. \label{fig6}}
\end{figure}

Now we shall apply the simulation results to the case 's' of Figures~\ref{fig1}c and \ref{fig2}, i.e.~to the time interval  12:50:55 to 12:51:15 to identify waves  responsible for the observed electron heating.
During this time interval the mean values of plasma parameters are: $v_{Te}=2000$ km/s, $r_e=4$ km, $r_p=600$ km, $f_{ce}=80$ Hz, $f_{lh}=2$ Hz, $B=3$ nT.
In Fig.~\ref{fig6} we show the $E_y$ component of the measured waves in the DSL coordinate system (close to GSE) decomposed into  discrete frequency dyads with orthogonal wavelets \citep{Mallat:1999}.    Orthogonality means that the time integral of the product of any pair of the frequency dyads is zero, and the decomposition is exact, i.e., the sum of all  components gives the original signal.

We note first in Fig.~\ref{fig6} that waves in channel 512 Hz have $\Omega=6.4$, so this channel and all higher frequencies are beyond the parameter region of bulk heating of electrons. They can provide only acceleration for the tail of the distribution.

Channels 32 - 256 Hz correspond to  (0.4 - 3.2) $f_{ce}$ so they are most likely ECD waves responsible for the observed stochastic electron heating with the leading role of the channel 256 Hz ($\Omega=3.2$) showing the largest amplitude, and providing the most efficient heating according to Fig. \ref{fig4}.

 The colormap  shows that the bulk heating extends up to $K\equiv k_x r_e\approx 10$, which implies that the bulk heating of the thermal population has to be done by waves $0.4f_{ce} \lesssim f<5f_{ce}$ with  wavelengths longer than

 \begin{equation}
 \lambda_K=2\pi r_e/K \approx 2.5\; \mathrm{km},
 \end{equation}
which means that  the electrostatic waves or solitary structures on the Debye length scales would not participate in the perpendicular bulk heating of  electrons. Lower hybrid waves, whistlers, and magnetosonic waves at $f \lesssim f_{lh}$ are also excluded.

An interesting question is why the ECD waves with a modest amplitude of 10 mV/m (Fig.~\ref{fig6}) produce $\chi_e\sim10$ and stochastic heating in the present case, while waves at amplitude 100 mV/m give only $\chi_e<1$ and quasi-adiabatic heating in perpendicular shocks \citep{Stasiewicz:2020c}. The answer is in the scaling $\chi \propto E B^{-2}$, because in this case $B\approx 3$ nT, while in perpendicular shocks $B\approx 30$ nT, which gives factor of 10$^{-2}$  that overrides factor of 10 in the $E$-field amplitude.

\subsection{Isotropization of the particle distribution} \label{seq_iso}

\begin{figure}
\includegraphics[width=7cm]{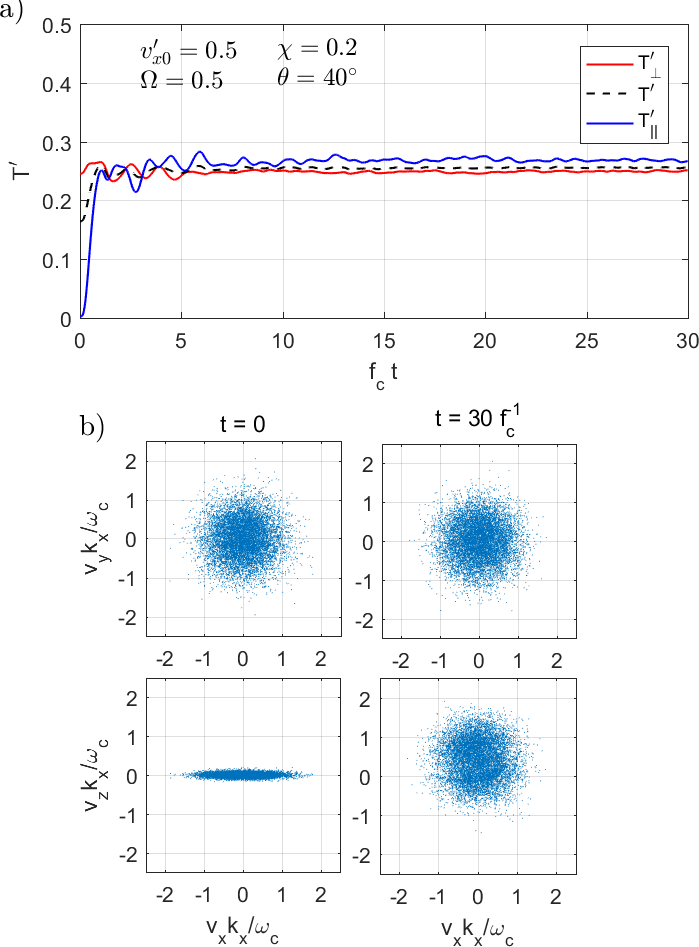}
\caption{Isotropization of electrons showing a) the time development of the kinetic temperatures, and b) the distribution function at initial time $t=0$ (left column) and final time $t=30\,f_c^{-1}$ (right column), projected in the $v_x$-$v_y$ plane (top row) and in the $v_x$-$v_z$ plane (bottom row), for $\theta=40^\circ$, $\chi=0.2$, $\Omega=0.5$ and $v_{x0}^\prime=0.5$ with the initial temperature anisotropy $\tau=T_{\parallel0}/T_{\perp0}=0.01$. The distribution function becomes isotropic within a few cyclotron periods. \label{fig_isotrop_theta_40}}
\end{figure}

\begin{figure}
\includegraphics[width=7cm]{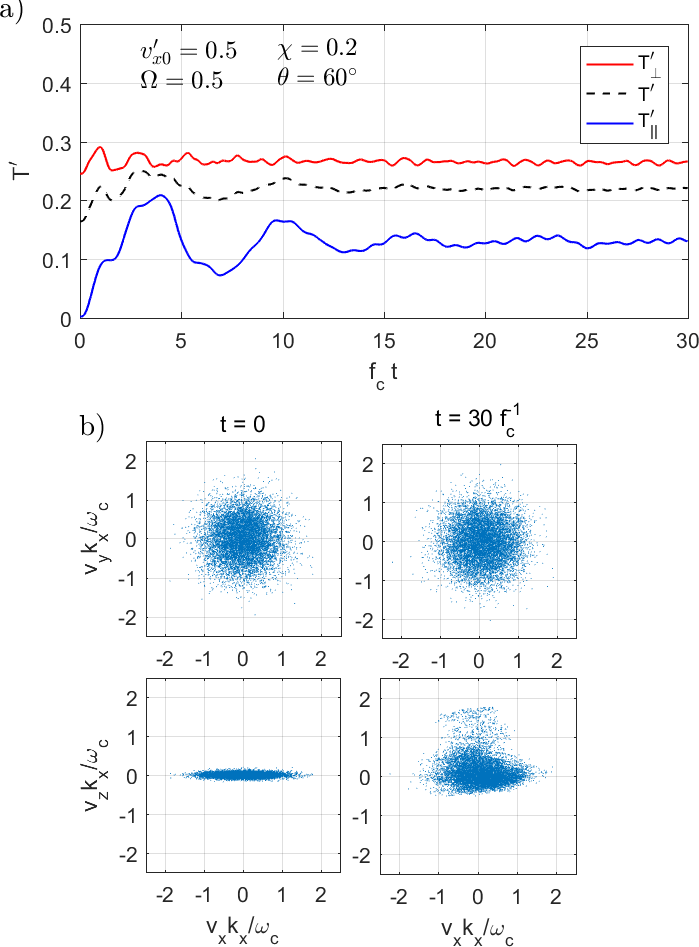}
\caption{Isotropization of electrons showing a) the time development of the kinetic temperatures, and b) the distribution function at initial time $t=0$ (left column) and final time $t=30\,f_c^{-1}$ (right column), projected in the $v_x$-$v_y$ plane (top row) and in the $v_x$-$v_z$ plane (bottom row), for $\theta=60^\circ$, $\chi=0.2$, $\Omega=0.5$ and $v_{x0}^\prime=0.5$ with the initial temperature anisotropy $\tau=T_{\parallel0}/T_{\perp0}=0.01$. The distribution function becomes only partially isotropic with a velocity tail of electrons along $v_z$. \label{fig_isotrop_theta_60}}
\end{figure}

\begin{figure*}
\centering
\includegraphics[width=15cm]{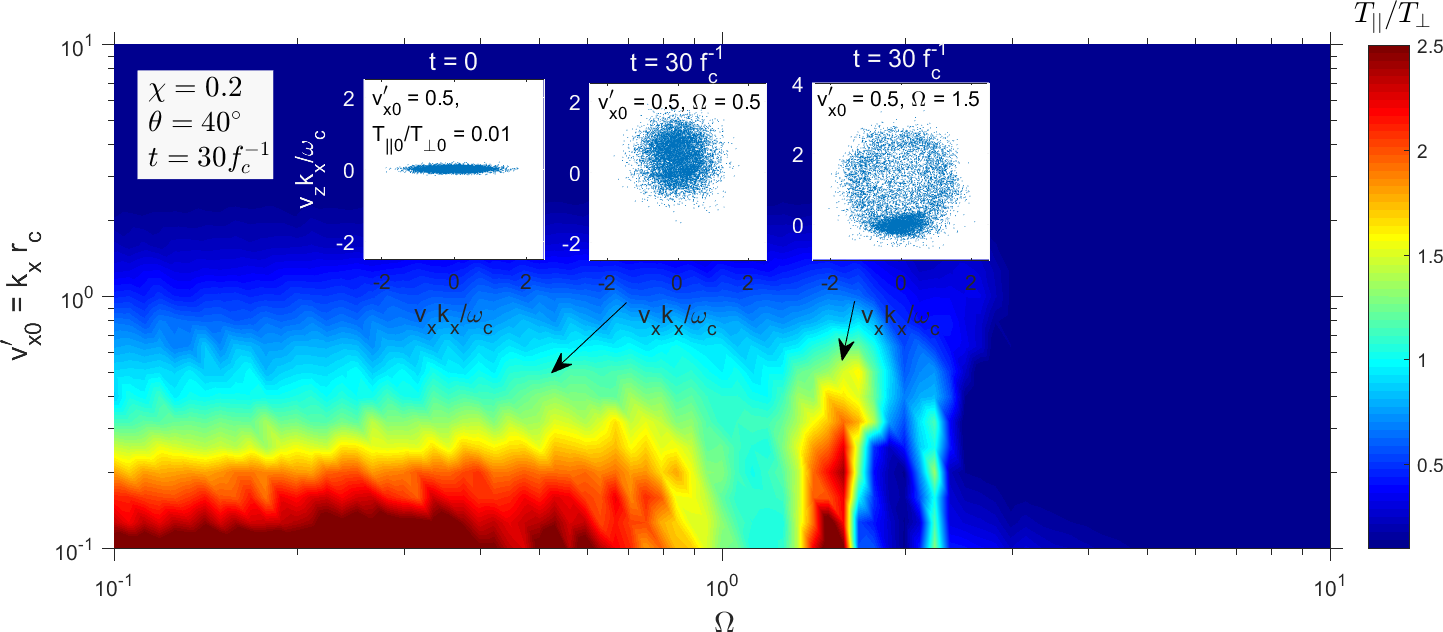}
\caption{Isotropization of electrons after 30 electron cyclotron periods by waves with normalized amplitude $\chi=0.2$ and angle $\theta=40^\circ$ to the magnetic field, and a range of values for $\Omega$ and $v_{x0}^\prime$ with the initial temperature anisotropy $\tau=T_{\parallel0}/T_{\perp0}=0.01$. The insets show the initial and final distribution functions for $v_{x0}^\prime=0.5$ and different values of $\Omega$. \label{fig_isotrop_ECD}}
\end{figure*}

In view of the observations in Figs.~\ref{fig1}c  that electron distributions are nearly isotropic at shocks, it is of interest to investigate which type of waves are most efficient for the isotropization.  We employ a set of test particle simulation to assess the efficiency of isotropization for obliquely propagating electrostatic waves. Figures \ref{fig_isotrop_theta_40} and \ref{fig_isotrop_theta_60} show the time evolutions of the kinetic temperatures as well as the initial and final distribution functions projected in the $v_x$-$v_y$ and $v_x$-$v_z$ planes subject to an electrostatic wave propagating respectively at angles $\theta=40^\circ$ and $60^\circ$ to the magnetic field, at a normalized frequency $\Omega=0.5$ and wave amplitude $\chi=0.2$, for an initial electron distribution function with $v_{x0}^\prime=0.5$ and an initial temperature anisotropy $\tau=T_{\parallel0}/T_{\perp0}=0.01$. For $\theta=40^\circ$ in Fig.~\ref{fig_isotrop_theta_40}a, the electron distribution rapidly becomes isotropic within a few electron cyclotron periods, and the resulting electron distribution function seen in Fig.~\ref{fig_isotrop_theta_40}b becomes almost isothermal in the $v_x$-$v_z$ plane. In contrast, for $\theta=60^\circ$ the electrons are only partially isotropized as seen in Fig.~\ref{fig_isotrop_theta_60}a, and the resulting distribution at $t=30\,f_c^{-1}$ in Fig.~\ref{fig_isotrop_theta_60}b is non-isothermal in the $v_x$-$v_z$ plane with a high velocity tail of electrons in the parallel $v_z$ direction. For both $\theta=40^\circ$ and $60^\circ$ the perpendicular velocities are not significantly affected as seen in the $v_x$-$v_y$ projections of the distribution functions in Figs.~\ref{fig_isotrop_theta_40}b and \ref{fig_isotrop_theta_60}b. In general, larger values of $\theta$ do not lead to isotropization but to resonant acceleration along $v_z$ for lower values of $\Omega$ and smaller $v_{x0}^\prime=k_x r_c$.

The efficiency of isotropization for a range of values of $\Omega$ and $v_{x0}^\prime$ is shown in Fig.~\ref{fig_isotrop_ECD}. Efficient isotropization within a time of $30$ cyclotron periods takes place  for $\Omega\lesssim 1$ and $v_{x0}^\prime\lesssim 1$, where the resulting electron distribution functions become essentially isotropic. For $1\lesssim\Omega\lesssim 3$ there are narrow regions where the electrons are resonantly accelerated parallel to the magnetic field lines so that $T_{\parallel}>T_{\perp}$. In this case the electrons have gained a positive mean drift and hence have produced a net current parallel to the magnetic field lines, reflecting the action of a single wave propagating obliquely in the positive direction to the magnetic field. Such currents could be cancelled if there are ECD waves propagating at both positive and negative directions to the magnetic field lines. Isotropization is not efficient for $\Omega>3$. In short, isotropization takes place for obliquely propagating electrostatic waves with frequencies near or below the electron cyclotron frequency, with wavelengths larger than the electron Larmor radius.
 We have run isotropization with $\chi=0.2$ to show that this process occurs for waves with lower amplitudes, and is not necessarily linked with the stochastic heating that requires $\chi>1$.

\section{Conclusions}
This research has shown that electrons are heated by two different mechanisms at the Earth's quasi-parallel shock: stochastic and quasi-adiabatic. We have performed detailed analysis of two heating cases identified as 's' stochastic, where the electron temperature was increased by 10 eV, and quasi-adiabatic case 'q' with temperature increase by 15 eV. The temperature increase was by a factor of two in both cases.

The  stochastic heating  preferably occurs at low magnetic fields, when the electron heating function $\chi_e = m_e q_e^{-1} B^{-2}\mathrm{div}(\mathbf{E}_\perp)$ exceeds unity and is accomplished by the perpendicular electric field  provided by the ECD instability. We have shown that the bulk heating is most likely done by waves in the frequency range $(0.4-5)f_{ce}$ with wavelengths $\lambda>2$ km, while the tail of the distribution can be accelerated by shorter waves with higher frequency, $f>5f_{ce}$. The simulations also indicate that stochastic heating in some cases leads to flat-topped distribution functions frequently observed in the vicinity of shocks in space plasma \citep{Feldman:1983,Lefebvre:2007}.

 A different heating mechanism takes place in regions where the magnetic field is being compressed, and $|\chi_e|<1$. In this case the electrons are heated through the conservation of the magnetic moment $\propto T_{e\perp}/B$ in a process identified by \citet{Stasiewicz:2020c} as quasi-adiabatic. The perpendicular kinetic energy gained is redistributed to the parallel direction by the scattering by waves, leading to an almost isotropic electron distribution, following the isotropic temperature relation $T/B=(T_0/B_0)(B_0/B)^{\alpha}$ where the value of $\alpha$ =1/3-2/3 depends on the physical processes involved. In the case of quasi-perpendicular shocks \citep{Stasiewicz:2020c} the kinetic energy of the electrons is conserved during the isotropization by  electrostatic fields, leading to $\alpha=1/3$.
  In the present case of quasi-parallel shocks a part of the perpendicular energy gain is converted to  waves causing the isotropization, and it is found that a larger $\alpha=2/3$ better fits the data.

 The observations have  been confirmed by test particle simulations showing rapid heating and isotropization of electrons by electrostatic waves propagating at different angles to the magnetic field. The isotropization is most efficient when electrostatic waves propagate at angles $\lesssim  60^\circ$ to the magnetic field.
 
 The results of this work support the shock heating scenario that starts with compression of the density, which via the initial diamagnetic current triggers consecutively three cross-field current driven instabilities: LHD $\rightarrow$ MTS $\rightarrow$ ECD, which produce stochastic heating of ions and electrons, in addition to a common quasi-adiabatic heating of electrons on compressions of $B$.

\acknowledgments
The authors thank members of the MMS mission for making available the data.
MMS science data is made available through the Science Data Center at
the Laboratory for Atmospheric and Space Physics (LASP) at the
University of Colorado, Boulder: https://lasp.colorado.edu/mms/sdc/public/.
B.E. acknowledges support from the EPSRC (UK), grants EP/R004773/1 and EP/M009386/1.
\software{Data analysis was performed using the IRFU-Matlab analysis package available at https://github.com/irfu/irfu-matlab.}

\bibliographystyle{aasjournal}

\end{document}